\documentstyle[12pt,epsfig,amsmath,amssymb,pstricks]{article}
\begin{document}

\title{\bf Imposing system-observer symmetry on the von Neumann description of measurement}

\author{{Chris Fields}\\ \\
{\it 528 Zinnia Court}\\
{\it Sonoma, CA 95476 USA}\\ \\
{fieldsres@gmail.com}}
\maketitle

\begin{abstract}
By imposing system-observer symmetry on the von Neumann description of measurement, it is shown that the quantum measurement problem is structurally equivalent to a familiar reverse-engineering problem: that of describing the behavior of an arbitrary physical device as algorithm instantiation.  It is suggested that this problem can at best be given a relational solution.
\end{abstract}


\section{Introduction}

The usual description of quantum measurement, the one made precise by von Neumann \cite{vonNeumann:32} and included, in one form or another, in virtually every textbook, involves an observer $\mathbf{O}$ who performs two operations on a quantum system $\mathbf{S}$: (1) a ``preparation'' operation that is generally only specified by specifying its outcome, a quantum state $|\mathbf{S}_{\mathit{s}} \rangle$ of $\mathbf{S}$ expressed as a superposition $|\mathbf{S}_{\mathit{s}} \rangle = \mathit{\sum_{i} \lambda_{i} |s_{i} \rangle}$ in some complete, orthonormal basis $\{ |s_{i} \rangle \}$ of the Hilbert space $\mathcal{H}_{\mathbf{S}}$ of $\mathbf{S}$; and (2) a ``measurement'' operation that is specified by specifying a Hermitian operator $M_{s}$ with eigenvalues $\alpha_{i}$ such that for each $i$, $M_{s} |\mathbf{S}_{\mathit{s}} \rangle = \mathit{\alpha_{i} |s_{i} \rangle}$ with a classical probability $P_{i} = | \lambda_{i} |^{2}$.  As emphasized by Bohr \cite{bohr:58} and many others, this usual description involves a choice on the part of the observer: the choice of which degrees of freedom of $\mathbf{S}$ to measure (e.g. position or momentum), and hence the choice of which basis $\{ |s_{i} \rangle \}$ to employ to describe $\mathcal{H}_{\mathbf{S}}$.  This choice determines, up to practical issues of laboratory implementation, both the preparation procedure that yields the $\{ |s_{i} \rangle \}$-specific ``ready'' state $|\mathbf{S}_{\mathit{s}} \rangle$ and the observable $M_{s}$ for which the $\{ |s_{i} \rangle \}$ are eigenstates.  The nature and implementation of this choice are generally left unspecified; indeed what ``choice'' amounts to is typically regarded as a philosophical or possibly neuropsychological question, not a physical question.

As is well known, the von Neumann description of measurement raises a conceptual problem, the ``measurement problem,'' or in more recent language, the problem of the ``emergence of classicality.''  This is not a formal problem; ``collapse'' to a specific eigenstate $\alpha_{i} |s_{i} \rangle$ is easily achieved with a projector.  It is rather a \textit{physical} problem: what \textit{physically implements} the formal operation represented by $M_{s}$?  The present paper approaches this problem via the fundamental symmetry of which Newton's third law is a special case: once asymmetrical approximations made for calculational convenience are set aside, any physical interaction between two physical systems must have equivalent descriptions from either system's point of view.  The action of $\mathbf{O}$ on $\mathbf{S}$ represented by $M_{s}$ must, therefore, also be describable as an action of $\mathbf{S}$ on $\mathbf{O}$, one that produces an outcome state of $\mathbf{O}$ characterized by the eigenvalue $\alpha_{i}$ and the classical probability $P_{i}$.  This outcome state of $\mathbf{O}$ corresponds, clearly, to $\mathbf{O}$ ``having the information'' that $\mathbf{S}$ is in the eigenstate $|s_{i} \rangle$.  Treating measurement as an operation by a quantum system on an observer removes the possibility of appeals to ill-defined notions of ``choice'' and offers the possibility of exposing implicit and potentially non-physical assumptions about ``observers.''  As discussed below, this approach also clarifies the physical presuppositions of asymmetrical approximations that are typically made when formally describing measurement, such as the statistical treatment of the environment in decoherence theory.

The paper begins by assuming that minimal, non-relativistic quantum theory is correct within its domain, applies universally, and provides a complete description of physical dynamics.  This fundamental assumption assures that it makes sense to talk about a universal Hilbert space $\mathcal{H}_{\mathbf{U}}$ and a universal quantum state $|\mathbf{U} \rangle$ that evolves according to a universal Schr\"odinger equation $(\partial / \partial t) | \mathbf{U} \rangle = - (\imath / \hbar) \mathit{H}_{\mathbf{U}} | \mathbf{U} \rangle$, where $H_{\mathbf{U}}$ is a deterministic universal Hamiltonian.  It rules out any objective non-unitary ``collapse'' of $| \mathbf{U} \rangle$; it also requires that observers be treated as physical systems and rules out any non-physical mechanism by which beliefs of or choices made by observers influence physical dynamics.  A category-theoretic formalism is employed to show that the usual quantum measurement problem can be described as a failure of commutativity between physical processes on the one hand and the tensor-product decomposition of $\mathcal{H}_{\mathbf{U}}$ on the other.  Using this same formalism, it is then shown that quantum measurement is structurally equivalent to the familiar reverse-engineering process of constructing a formal semantics that interprets the behavior of an otherwise-uncharacterized physical device as an execution trace of either a classical or a quantum algorithm.  This structural equivalence suggests that ``measurement'' can be self-consistently viewed as a purely virtual process implemented by $H_{\mathbf{U}}$.  Adopting this view enables a formal representation of measurement in which physical and virtual processes commute, and hence a formal resolution of the measurement problem.  It highlights the deep question of how $H_{\mathbf{U}}$ physically implements the familiar classical semantics of separable ``systems,'' and suggests that this question can at best be given a relational answer.

\section{The measurement problem as a problem of operator commutativity}

Describing a sequence of physical events in terms of system preparation and measurement requires introducing a tensor-product structure (TPS) on $\mathcal{H}_{\mathbf{U}}$ that explicitly divides the physical degrees of freedom of $\mathbf{U}$ into (at least) three subsets: a set of degrees of freedom of the system of interest $\mathbf{S}$, a set of degrees of freedom of the observer $\mathbf{O}$, and a set comprising all degrees of freedom not assigned to either $\mathbf{S}$ or $\mathbf{O}$, which are conventionally considered to characterize the ``environment'' (or ``shared environment'') $\mathbf{E}$.  This environment can be taken to include any macroscopic apparatus that $\mathbf{O}$ employs to investigate $\mathbf{S}$.  It is standard to write this TPS as a simple equality $\mathcal{H}_{\mathbf{S}} \otimes \mathcal{H}_{\mathbf{E}} \otimes \mathcal{H}_{\mathbf{O}} = \mathcal{H}_{\mathbf{U}}$.  Zurek, for example, adopts as ``axiom(o)'' of quantum theory that ``the Universe consists of systems'' and ``a composite system can be described by a tensor product of the Hilbert spaces of the constituent systems'' (\cite{zurek:03} p. 746; see also \cite{zurek:05env}, p. 2; \cite{zurek:07}, p. 3) in order to justify this practice.  

From an information-theoretic perspective, imposing a TPS on $\mathcal{H}_{\mathbf{U}}$ is an \textit{operation}, one that partitions the degrees of freedom of $\mathbf{U}$.  Thus while the evolution of $\mathbf{U}$ as a whole can be discussed within the category $(\mathcal{H}, \mathit{A})$ of Hilbert spaces $\mathcal{H}$ and Hilbert-space automorphisms $A$, the imposition of a TPS requires the richer category $(\mathcal{H}, \mathit{A}, \otimes)$ that includes Hilbert-space tensor products as additional morphisms.  Let $\mathcal{D} : (\mathcal{H}, \mathit{A}) \rightarrow (\mathcal{H}, \mathit{A}, \otimes)$ be the \textit{decomposition} functor that introduces $\otimes$ and hence TPSs into $(\mathcal{H}, \mathit{A})$, and let $\mathcal{D}_{\mathbf{SO}} : \mathcal{H}_{\mathbf{U}} \mapsto \mathcal{H}_{\mathbf{S}} \otimes \mathcal{H}_{\mathbf{E}} \otimes \mathcal{H}_{\mathbf{O}}$ be a particular decomposition operator acting on the particular Hilbert space $\mathcal{H}_{\mathbf{U}}$.  This explicit representation of decomposition allows the explicit statement of an assumed symmetry of the physical world that is left implicit within the standard formalism: for any specified $\mathbf{S}$ and $\mathbf{O}$, the decomposition mapping $\mathcal{D}_{\mathbf{SO}} = \mathcal{I}_{\mathbf{U}}$, where $\mathcal{I}_{\mathbf{U}}$ is the Identity operator on $\mathcal{H}_{\mathbf{U}}$.  This symmetry, called ``decompositional equivalence'' in \cite{fields:12a, fields:12b, fields:12c}, is what allows $\mathcal{H}_{\mathbf{S}} \otimes \mathcal{H}_{\mathbf{E}} \otimes \mathcal{H}_{\mathbf{O}} = \mathcal{H}_{\mathbf{U}}$ to be written as an identity, and what allows $H_{\mathbf{U}}$ to be written as a sum $H_{\mathbf{U}} = \mathit{\sum_{ij} H_{ij}}$ of Hamiltonians $H_{ij}$ coupling individual degrees of freedom $i$ and $j$.  It states that decompositions into systems have no effect on the state $|\mathbf{U} \rangle$ or the Hamiltonian $H_{\mathbf{U}}$; different decompositions enable different \textit{descriptions} of the physics of $\mathbf{U}$, but do not \textit{change} of the physics of $\mathbf{U}$.  The assumption of decompositional equivalence is clearly essential to the practice of science; without it, alternative descriptions cannot be regarded as descriptions of the same universe.

Using this explicit representation of decomposition, the usual von Neumann description of measurement can be represented in diagrammatic form as in Fig. 1.  Here the universal state $|\mathbf{U} \rangle$ evolves under the action of the universal propagator $e^{-(i/ \hbar)H_{\mathbf{U}}(t)}$ in an entirely decomposition-independent way.  The decomposition operator $\mathcal{D}_{\mathbf{SO}}$ acts at the time points $t_{0}$, $t_{1}$ and $t_{2}$ to construct, at each time point $t_{n}$, a TPS $\mathcal{H}_{\mathbf{S}}\mathit{(t_{n})} \otimes \mathcal{H}_{\mathbf{E}}\mathit{(t_{n})} \otimes \mathcal{H}_{\mathbf{O}}\mathit{(t_{n})}$.  It is an assumption, often implicit, of the von Neumann representation of measurement that this TPS is time-invariant, i.e. that no degrees of freedom are exchanged between partitions defined by the TPS; the time labels are kept explicit in Fig. 1 for consistency with later diagrams.  The evolution of the states of this assumed-to-be time-invariant TPS is given by the standard von Neumann chain.  Between $t_{0}$ and $t_{1}$, a preparation operation acts on an unspecified initial state $|\mathbf{S} (\mathit{t_{0}})\rangle \otimes |\mathbf{E} (\mathit{t_{0}})\rangle \otimes |\mathbf{O} (\mathit{t_{0}})\rangle$ to yield the prepared state $\mathit{\sum_{i} \lambda_{i} |s_{i} \rangle} \otimes |\mathbf{E} (\mathit{t_{1}})\rangle \otimes |\mathbf{O}_{\mathit{s}} \rangle$, where the notation `$|\mathbf{O}_{\mathit{s}} \rangle$' is employed to indicate that at $t_{1}$, the observer $\mathbf{O}$ has the information that the system $\mathbf{S}$ has been prepared in the state $\sum_{i} \lambda_{i} |s_{i} \rangle$.  Between $t_{1}$ and $t_{2}$ the Hermitian operator $M_{s}$ acts on the prepared system state to yield, with probability $P_{i}$, the eigenstate $\alpha_{i} |s_{i} \rangle$; the notation `$|\mathbf{O}_{\mathit{i}} \rangle$' indicates that at $t_{2}$, $\mathbf{O}$ has the information that $\mathbf{S}$ occupies this eigenstate.

\psset{xunit=1cm,yunit=1cm}
\begin{pspicture}(0,0)(16,7.5)

\put(0.5,6.5){$|\mathbf{S} (\mathit{t_{0}})\rangle \otimes |\mathbf{E} (\mathit{t_{0}})\rangle \otimes |\mathbf{O} (\mathit{t_{0}})\rangle$}
\put(5.7,6.5){$\mapsto$}
\put(6.3,6.5){$\mathit{\sum_{i} \lambda_{i} |s_{i} \rangle} \otimes |\mathbf{E} (\mathit{t_{1}})\rangle \otimes |\mathbf{O}_{\mathit{s}} \rangle$}
\put(11.3,6.5){$\mapsto$}
\put(12,6.5){$\alpha_{i} |s_{i} \rangle \otimes |\mathbf{E} (\mathit{t_{2}})\rangle \otimes |\mathbf{O}_{\mathit{i}} \rangle$}

\put(4.7,5.9){\textit{Preparation}}
\put(11.3,5.9){$M_{s}$}

\put(.2,5.5){$\mathcal{H}_{\mathbf{S}}\mathit{(t_{0})} \otimes \mathcal{H}_{\mathbf{E}}\mathit{(t_{0})} \otimes \mathcal{H}_{\mathbf{O}}\mathit{(t_{0})}$}
\put(5.2,5.6){\vector(1,0){.8}}
\put(6.2,5.5){$\mathcal{H}_{\mathbf{S}}\mathit{(t_{1})} \otimes \mathcal{H}_{\mathbf{E}}\mathit{(t_{1})} \otimes \mathcal{H}_{\mathbf{O}}\mathit{(t_{1})}$}
\put(11.2,5.6){\vector(1,0){.8}}
\put(12.2,5.5){$\mathcal{H}_{\mathbf{S}}\mathit{(t_{2})} \otimes \mathcal{H}_{\mathbf{E}}\mathit{(t_{2})} \otimes \mathcal{H}_{\mathbf{O}}\mathit{(t_{2})}$}

\put(1.3,4.3){$\mathcal{D}_{\mathbf{SO}} (\mathit{t_{0}})$}
\put(3,3.5){\vector(0,1){1.5}}
\put(6.8,4.3){$\mathcal{D}_{\mathbf{SO}} (\mathit{t_{1}})$}
\put(8.5,3.5){\vector(0,1){1.5}}
\put(12.3,4.3){$\mathcal{D}_{\mathbf{SO}} (\mathit{t_{2}})$}
\put(13.9,3.5){\vector(0,1){1.5}}

\put(4.5,3.3){$e^{-(i/ \hbar)H_{\mathbf{U}}(t)} |_{t_{0} \rightarrow t_{1}}$}
\put(2.7,3){$\mathcal{H}_{\mathbf{U}}$}
\put(4,3.1){\vector(1,0){3.5}}
\put(8.2,3){$\mathcal{H}_{\mathbf{U}}$}
\put(9.5,3.1){\vector(1,0){3.5}}
\put(13.6,3){$\mathcal{H}_{\mathbf{U}}$}
\put(10,3.3){$e^{-(i/ \hbar)H_{\mathbf{U}}(t)} |_{t_{1} \rightarrow t_{2}}$}

\put(0.5,1.5){\textit{Fig. 1:}  The usual von Neumann representation of measurement, with the decomposition}
\put(0.5,1){mapping $\mathcal{D}_{\mathbf{SO}}$ made explicit.  The mapping $M_{s}$ is a Hermitian observable associated with}
\put(0.5,0.5){the basis $\{ |s_{i} \rangle \}$.  The upper row of mappings is the familiar von Neumann chain.}
\end{pspicture}

The usual quantum measurement problem can be stated succinctly in this representation: the diagram shown in Fig. 1 does not commute, i.e. $\mathcal{D}_{\mathbf{SO}} (\mathit{t_{2}}) \circ e^{-(i/ \hbar)H_{\mathbf{U}}(t)} |_{t_{0} \rightarrow t_{2}} \neq \mathit{M_{s} \circ ~Preparation~} \circ \mathcal{D}_{\mathbf{SO}} (\mathit{t_{0}})$ where `$\circ$' indicates composition of mappings and is read ``following.''  In particular, $\alpha_{i} |s_{i} \rangle$ is a ``collapsed'' eigenstate of $M_{s}$, while nothing has collapsed in $|\mathbf{U} \rangle$.  This failure of commutativity makes it impossible to say in a well-defined way how the evolution of $|\mathbf{U} \rangle$ physically \textit{implements} the measurement process, as it must do if quantum theory is to be complete.

As is well known, different physical interpretations of quantum theory deal with this failure of commutativity in different ways.  Non-ontic approaches, including the Copenhagen interpretation if it is viewed purely epistemically, reject the identification of the states on the top line as \textit{physical} states, treating them instead as informational states in the mind of the observer (e.g. \cite{clifton:03, fuchs:10, chiribella:11}).  Such approaches to date make no attempt to explain \textit{how} information about physical states gets into the mind of the observer.  Fuchs \cite{fuchs:10}, for example, explicitly rejects this question as outside the domain of the theory, thus implicitly rejecting either the physicality of observers or the universality of quantum theory as a description of physical processes.  The Everett (e.g. \cite{everett:57, wallace:10, tegmark:10}) and decoherent histories (e.g. \cite{griffiths:02, hartle:08}) approaches expand the top line into the set of all possible mutually-decoherent top lines (i.e. for all possible values of $i$), and treat all as equally physical and hence equally ontic.  Classicality is taken to be imposed by decoherence; both system and observer \cite{tegmark:10} or the observer's mind \cite{zeh:00} are taken to exist in multiple ``copies'' on different, mutually-inaccessible classical branches or histories.  In order for these multiple copies to be \textit{identified} as physical and hence quantum systems, i.e. as the TPS components $\mathbf{S}$ or $\mathbf{O}$ in Fig. 1, the TPS $\mathcal{H}_{\mathbf{S}} \otimes \mathcal{H}_{\mathbf{E}} \otimes \mathcal{H}_{\mathbf{O}}$ must be assumed to be invariant across branches or histories.  The environment as witness formulation of decoherence theory \cite{zurek:04, zurek:05} replaces the top line with separate local system-environment and environment-observer interactions, and argues that observers can only extract from their local environments information about eigenstates of the local system-environment interaction.  As these eigenstates are selected and redundantly encoded by the environment independently of the observer, the local environment-observer interaction and the observer's information are both effectively classical regardless of their physical implementation.  Whether effective classicality is attributed within these various approaches to records stored in the memories of automated systems such as laboratory computers (i.e. in the environment $\mathbf{E}$) as well as to records stored in the memories of human or other animate observers is rarely discussed explicitly and seldom made clear.  Recent surveys of attendees at quantum foundations conferences indicate little consensus on the most basic questions regarding the physical interpretation of quantum theory \cite{schloss:13, sommer:13}; such fundamental disagreement about the physical interpretation of the theory constitutes \textit{prima facie} evidence that the meanings of the formal expressions employed by the theory require clarification.

One relevant issue that standard interpretations of quantum theory, and hence standard explanations of the emergence of classicality universally neglect is that the decomposition operator $\mathcal{D}_{\mathbf{SO}}$ by itself introduces classicality into Fig. 1, quite independently of assumptions or interpretations regarding collapse or decoherence.  By imposing a TPS on $\mathcal{H}_{\mathbf{U}}$, $\mathcal{D}_{\mathbf{SO}}$ allocates, either explicitly or implicitly, every degree of freedom of $\mathcal{H}_{\mathbf{U}}$ to one component or another of the TPS.  In some cases, this allocation of degrees of freedom is physically paradoxical, as for example when a collection of microscopic degrees of freedom is considered to be part of the environment of a macroscopic degree of freedom - such as the center-of-mass position of a bulk material object - that exists only as a notional mathematical consequence of its microscopic ``environment'' (e.g. \cite{omnes:92} p. 354 \textit{ff}).  Even in cases that are not paradoxical in this way, however, how the components of a TPS evolving under the action of $e^{-(i/ \hbar)H_{\mathbf{U}}(t)}$ relate to our ordinary sense of an ``object'' or its ``environment'' is far from clear.  For example, consider an Ag ion passing through a Stern-Gerlach apparatus.  From the simplified formal perspective of a typical textbook, the ion's spin is the system of interest $\mathbf{S}$; heating the ion in the ion source prepares (for example) the state $| \mathbf{S}_{\mathit{z}} \rangle = (1/\sqrt(2) (| \uparrow \rangle + | \downarrow \rangle)$.  From a physical perspective, however, it is the ion itself that is heated and emitted from the source and that passes through the apparatus, and it is the ion's position that is eventually detected, via a physical interaction with its electric charge.  Hence physically, it is the ion itself, as a package including all of its degrees of freedom, that $\mathcal{D}_{\mathbf{SO}}$ allocates to the TPS component $\mathbf{S}$.  Consistent with von Neuman's representation, no one considers the ion being ``part of $\mathbf{S}$'' as one possible value of a physical (i.e. quantum) degree of freedom; no one writes $|\psi_{ion} \rangle = (1/\sqrt{2}) (|\mathit{part ~of } ~\mathbf{S} \rangle + |\mathit{part ~of } ~\mathbf{E} \rangle)$.  All acknowledge, however, that prior to the experiment the ion was part of the ion source, and that by the end of the measurement the ion has been irreversibly absorbed by the detector; in both of these conditions it is ``part of $\mathbf{E}$'' in every meaningful sense.  This is clearly problematic, as it implies that $\mathcal{H}_{\mathbf{E}}(\mathit{t_{0}}) \neq \mathcal{H}_{\mathbf{E}} (\mathit{t_{1}}) \neq \mathcal{H}_{\mathbf{E}} (\mathit{t_{2}})$, and in particular that both $Preparation$ and $M_{s}$ alter the TPS of $\mathcal{H}_{\mathbf{U}}$.  Any such change violates von Neumann's assumption of a fixed, time-invariant decomposition, and hence threatens the self-consistency of the von Neumann representation of measurement.  On a deeper level, the conceptual conflict between the mathematical construction of a fixed TPS and the experimental generation of ``systems'' from the bulk properties of the world suggests that the physical separability implied by a TPS may be a poorly-understood approximation even under the best of circumstances.  It suggests, in other words, that the contextuality demonstrated by the Kochen-Specker and similar theorems \cite{kochen:67, mermin:93} may a more ubiquitous feature of physical reality than is generally appreciated.

Lack of clarity about what decomposition of $\mathcal{H}_{\mathbf{U}}$ into a TPS means physically is especially profound in formal approaches that appeal to environmental decoherence during the measurement process.  Decoherence depends for its definition on a TPS; it depends for physical effectiveness on the assumption that $\mathbf{E}$ is large with respect to both $\mathbf{S}$ and $\mathbf{O}$, an assumption that is often rendered epistemically as an assumption that $\mathbf{O}$ does not (or cannot) measure the quantum state $| \mathbf{E} \rangle$, and rendered formally by treating $\mathbf{E}$ as a heat bath.  If the system $\mathbf{S}$ must be \textit{produced from} $\mathbf{E}$, however, not just observation but specific, state-dependent manipulation of $\mathbf{E}$ by $\mathbf{O}$ is required.  Such observations and manipulations can be dismissed as ``classical,'' but not in a theoretical context intended to explain the ``emergence of classicality'' via decoherence.  The mechanism of decoherence depends, moreover, on system-environment entanglement and hence on the progressive blurring of any physical distinction between ``system'' and ``environment.''  Indeed if decompositional equivalence is true in $\mathcal{H}_{\mathbf{U}}$, as it must be for quantum theory to be even approximately accurate, the operation $\mathcal{D}_{\mathbf{SO}}$ can have no \textit{physical} consequences; in this case what it means to assume separability and call $|\mathbf{S} \rangle$, $|\mathbf{E} \rangle$ or $|\mathbf{O} \rangle$ \textit{physical states} is itself unclear.  Reversing the usual perspective on measurement and asking how an observer, as opposed to an observed system, gets into his, her or its post-measurement state serves to isolate and emphasize the role of such decomposition-imposed classicality in creating the measurement problem.

\section{Measurement as an action of a system on an observer}

If one follows Landauer's \textit{dictum} that all information is at all times physically encoded \cite{landauer:99}, then the states $|\mathbf{O}_{\mathit{s}} \rangle$ and $|\mathbf{O}_{\mathit{i}} \rangle$ must be \textit{physical} states of the observer that encode the information ``the system is now in the state $\sum_{i} \lambda_{i} |s_{i} \rangle$'' and ``the system is now in the state $\alpha_{i} |s_{i} \rangle$'' respectively.  Although these items of information \textit{describe} quantum states of $\mathbf{S}$, they are themselves classical: they may be expressed using finite strings of fixed symbols, and written on a classical memory such as a printed page.  It is, therefore, generally supposed that they are encoded by classical states of the observer, i.e. that $|\mathbf{O}_{\mathit{s}} \rangle$ and $|\mathbf{O}_{\mathit{i}} \rangle$ are classical - or ``effectively classical'' - physical states.  The point of both the environment as witness formulation of decoherence and the ``quantum consciousness'' theories of von Neumann \cite{vonNeumann:32}, Wigner \cite{wigner:62}, or more recently Hameroff and Penrose \cite{hameroff:96} or Stapp \cite{stapp:01} is to show \textit{how} such physical states of the observer could become classical.

Before asking how $|\mathbf{O}_{\mathit{s}} \rangle$ and $|\mathbf{O}_{\mathit{i}} \rangle$ could be or even appear to be classical states, however, it is reasonable to ask how $\mathbf{O}$ gets into either of these states.  The answer can only be: by \textit{physically interacting} with $\mathbf{S}$, or in the environment as witness formulation, with $\mathbf{E}$.  As a first step toward describing this interaction, let us define a mapping $\mathcal{R} : \mathit{A \rightarrow A}$ on automorphisms $A$ of $\mathbf{U}$ such that $\mathcal{R}(\mathit{M_{s}})$ is an Hermitian operator representing the physical action of $\mathbf{S}$ on $\mathbf{O}$.  The measurement process can then be described as in Fig. 2, where as in Fig. 1 the top line represents the von Neumann chain.  Here the notation `$|\mathbf{S}_{\mathit{o}} \rangle$' indicates the state of $\mathbf{S}$ after ``preparing'' $\mathbf{O}$ in the basis $\{ |o_{i} \rangle \}$ and the notation `$|\mathbf{S}_{\mathit{i}} \rangle$' indicates the state of $\mathbf{S}$ after interacting with $\mathbf{O}$ in a way that places $\mathbf{O}$ in the eigenstate $\alpha_{i} |o_{i} \rangle$ with probability $P_{i}$.

\psset{xunit=1cm,yunit=1cm}
\begin{pspicture}(0,0)(16,7.5)

\put(0.5,6.5){$|\mathbf{S} (\mathit{t_{0}})\rangle \otimes |\mathbf{E} (\mathit{t_{0}})\rangle \otimes |\mathbf{O} (\mathit{t_{0}})\rangle$}
\put(5.7,6.5){$\mapsto$}
\put(6.3,6.5){$|\mathbf{S}_{\mathit{o}} \rangle \otimes |\mathbf{E} (\mathit{t_{1}})\rangle \otimes \mathit{\sum_{i} \lambda_{i} |o_{i} \rangle}$}
\put(11.3,6.5){$\mapsto$}
\put(12,6.5){$|\mathbf{S}_{\mathit{i}} \rangle \otimes |\mathbf{E} (\mathit{t_{2}})\rangle \otimes \alpha_{i} |o_{i} \rangle$}

\put(4.5,5.9){$\mathcal{R}$(\textit{Preparation})}
\put(11,5.9){$\mathcal{R} \mathit{(M_{s})}$}

\put(.2,5.5){$\mathcal{H}_{\mathbf{S}}\mathit{(t_{0})} \otimes \mathcal{H}_{\mathbf{E}}\mathit{(t_{0})} \otimes \mathcal{H}_{\mathbf{O}}\mathit{(t_{0})}$}
\put(5.2,5.6){\vector(1,0){.8}}
\put(6.2,5.5){$\mathcal{H}_{\mathbf{S}}\mathit{(t_{1})} \otimes \mathcal{H}_{\mathbf{E}}\mathit{(t_{1})} \otimes \mathcal{H}_{\mathbf{O}}\mathit{(t_{1})}$}
\put(11.2,5.6){\vector(1,0){.8}}
\put(12.2,5.5){$\mathcal{H}_{\mathbf{S}}\mathit{(t_{2})} \otimes \mathcal{H}_{\mathbf{E}}\mathit{(t_{2})} \otimes \mathcal{H}_{\mathbf{O}}\mathit{(t_{2})}$}

\put(1.3,4.3){$\mathcal{D}_{\mathbf{SO}} (\mathit{t_{0}})$}
\put(3,3.5){\vector(0,1){1.5}}
\put(6.8,4.3){$\mathcal{D}_{\mathbf{SO}} (\mathit{t_{1}})$}
\put(8.5,3.5){\vector(0,1){1.5}}
\put(12.3,4.3){$\mathcal{D}_{\mathbf{SO}} (\mathit{t_{2}})$}
\put(13.9,3.5){\vector(0,1){1.5}}

\put(4.5,3.3){$e^{-(i/ \hbar)H_{\mathbf{U}}(t)} |_{t_{0} \rightarrow t_{1}}$}
\put(2.7,3){$\mathcal{H}_{\mathbf{U}}$}
\put(4,3.1){\vector(1,0){3.5}}
\put(8.2,3){$\mathcal{H}_{\mathbf{U}}$}
\put(9.5,3.1){\vector(1,0){3.5}}
\put(13.6,3){$\mathcal{H}_{\mathbf{U}}$}
\put(10,3.3){$e^{-(i/ \hbar)H_{\mathbf{U}}(t)} |_{t_{1} \rightarrow t_{2}}$}

\put(0.5,1.5){\textit{Fig. 2:}  Measurement represented as an action of the system $\mathbf{S}$ on the observer $\mathbf{O}$.}
\put(0.5,1){The operator $\mathcal{R}$ transforms $M_{s}$ into the action $\mathcal{R}(\mathit{M_{s}})$ on $|\mathbf{S} \rangle \otimes |\mathbf{O} \rangle$.  The upper row}
\put(0.5,0.5){of mappings is the transformed von Neumann chain.}
\end{pspicture}

The reversed von Neumann chain shown in Fig. 2 allows the question of the role of decoherence in measurement to be asked independently of any implicit and potentially non-physical assumptions about observers.  In Fig. 2, decoherence does not act on $\mathbf{S}$ to einselect an eigenstate; it acts on $\mathbf{O}$, and einselects eigenstates in the $\{ |o_{i} \rangle \}$ basis.  These eigenstates of $\mathbf{O}$ must, therefore, be eigenstates of the $\mathbf{S}$-$\mathbf{O}$ interaction, or in the environment as witness formulation, of the $\mathbf{E}$-$\mathbf{O}$ interaction.  This interaction is the interaction that physically implements $\mathcal{R}(\mathit{M_{s}})$.  Hence in the reversed von Neumann chain, measurement and einselection are the same physical action on the observer.  Einselection and hence decoherence cannot, therefore, play any role in the measurement process that is not already played by $\mathcal{R}(\mathit{M_{s}})$; in particular, neither decoherence nor einselection can be \textit{prerequisites} of $\mathcal{R}(\mathit{M_{s}})$, as they are in standard presentations of the environment as witness formulation.  Decoherence cannot, in other words, create either physical or effective classicality that is not already created by the reversed ``measurement'' operation performed by $\mathbf{S}$, or in the environment as witness formulation by $\mathbf{E}$, on the observer $\mathbf{O}$.

To emphasize this point, let us adopt the perspective of the environoment as witness, with its notion that the observer $\mathbf{O}$ is, as in the formalism of \cite{zurek:04, zurek:05}, restricted to a ``fragment'' $\mathbf{F}$ of the environment $\mathbf{E}$ that is sufficiently distant from the system $\mathbf{S}$ that the local $\mathbf{E}$-$\mathbf{O}$ interaction - i.e. the $\mathbf{F}$-$\mathbf{O}$ interaction - can be taken to have no effect on the $\mathbf{S}$-$\mathbf{E}$ interaction and hence no effect on the state of $\mathbf{S}$.  In this situation, $\mathbf{O}$ interacts with and extracts information from encodings of the distant system state $|\mathbf{S} \rangle$ in the local state $|\mathbf{F} \rangle$ of the environmental fragment $\mathbf{F}$; the local encoding of the position of a distant system by the ambient visual-spectrum photon field near the observer is the canonical example \cite{zurek:06, zurek:09}.  From the present, reversed perspective, the action of $\mathbf{F}$ on $\mathbf{O}$ implements this information transfer.  One can now ask: how does the action of $\mathbf{F}$ on $\mathbf{O}$ transfer information about the distant system $\mathbf{S}$, and in particular, how does this action result in $\mathbf{O}$ occupying $\alpha_{i} |o_{i} \rangle$ with probability $P_{i}$?  What is it about $|\mathbf{F} \rangle$ that makes it a \textit{specific} encoding of $|\mathbf{S} \rangle$?  In a universe satisfying decompositional equivalence, this question clearly can have no answer.  Nothing makes $|\mathbf{F} \rangle$ specific to $|\mathbf{S} \rangle$; indeed $|\mathbf{F} \rangle$ equally encodes the states of every collection of degrees of freedom outside of $\mathbf{F}$ \cite{fields:12b, fields:12c}.  The action of $\mathbf{F}$ on $\mathbf{O}$ cannot, therefore, place $\mathbf{O}$ in the $\mathbf{S}$-specific state $\alpha_{i} |o_{i} \rangle$.  If $\mathbf{O}$ is restricted to a distant environmental fragment $\mathbf{F}$, Fig. 2 not only fails to commute; the operator $\mathcal{R}(\mathit{M_{s}})$ becomes physically undefined.

While this analysis of the reversed measurement situation illustrated by Fig. 2 shows that decoherence cannot explain the effective classicality of measurement outcomes, it fails to show what does explain it.  Like Fig. 1, Fig. 2 does not commute; nothing in $|\mathbf{U}\rangle$ corresponds to the ``collapsed'' eigenstate $\alpha_{i} |o_{i}\rangle$.  The primary problem with Fig. 2, however, is that the physical basis states $|o_{i} \rangle$ remain uncharacterized.  All that is known is the classical information that these basis states encode: $|o_{i} \rangle$ encodes the classical information `$\mathbf{S}$ is in $|s_{i} \rangle$'.  Any reflection on the physical structure of human observers - or even of general-purpose computers employed as data recorders - moreover suggests that the physical states $|o_{i} \rangle$ may differ almost arbitrarily between observers.  If this is the case, the operator $\mathcal{R}(\mathit{M_{s}})$ once again becomes physically undefined; $\mathbf{S}$ must act in different ways on different observers to produce the same encoded information as an outcome.  If $\mathcal{R}(\mathit{M_{s}})$ is not well-defined across multiple observers, however, the unreversed measurement operator $M_{s}$ cannot be well-defined across multiple observers either.  We write down formal expressions such as `$\hat{p} = - \imath \hbar (\partial / \partial x)$' that \textit{appear to be} well-defined, but these expressions clearly do not describe the physical interaction of a human being (or a physically-implemented computer) and a physical system.  Indeed, the formal expression `$\hat{p} = - \imath \hbar (\partial / \partial x)$' does not even reference the Hilbert space over which it is meant to be defined; it is simply assumed in ordinary practice that $\hat{p} = - \imath \hbar (\partial / \partial x)$ acts on whatever Hilbert space is of interest.  We have no idea, in other words, how $\hat{p} = - \imath \hbar (\partial / \partial x)$ is \textit{implemented} by $H_{\mathbf{U}}$, or for that matter by the local interaction $H_{\mathbf{S} \text{-} \mathbf{O}}$ between any designated $\mathbf{S}$ and $\mathbf{O}$.

\section{Representing encoded classical information explicitly}

What is known about the physical state $\alpha_{i} |o_{i} \rangle$ of the observer is that it represents the classical information `$\mathbf{S}$ is in $|s_{i} \rangle$' and has a classical probability $P_{i}$ as an outcome of the to-be-defined physical action $\mathcal{R}(\mathit{M_{s}})$.  Abstract ``states'' comprising encoded classical information are familiar entities: states of classical computational processes are such things.  Let us define a functor $\mathcal{V} : (\mathcal{H}, \mathit{A}) \rightarrow (\mathcal{M}, \twoheadrightarrow)$ that maps the category $(\mathcal{H}, \mathit{A})$ of Hilbert spaces and Hilbert-space automorphisms to the category $(\mathcal{M}, \twoheadrightarrow)$ of abstract virtual machines \cite{goldberg:74, tan:76, hopcroft:79} and virtual-machine state transitions.  Let $\mathcal{V}_{\mathbf{O}}(\mathit{t_{n}})$ be an instance of $\mathcal{V}$ acting on the particular Hilbert space $\mathcal{H}_{\mathbf{O}}(\mathit{t_{n}})$ to generate the virtual-machine state implemented by $\mathbf{O}$ at $t_{n}$.  The diagram shown in Fig. 2 can then be extended to Fig. 3, using the notation `[$X$]' to indicate the virtual machine state encoding the classical information `$X$'.

\psset{xunit=1cm,yunit=1cm}
\begin{pspicture}(0,0)(16,10.5)

\put(3.5,9.4){$\mathcal{V}_{\mathbf{O}}$(\textit{Preparation})}
\put(11,9.4){$\mathcal{V}_{\mathbf{O}}(\mathit{M_{s}})$}

\put(2.7,9){[$\mathbf{S}$]}
\psline{->>}(4.2,9.1)(5.7,9.1)
\put(6.5,9){[$\mathbf{S}$ \textit{prepared in} $\{ |s_{i} \rangle \}$]}
\psline{->>}(11,9.1)(12.5,9.1)
\put(13,9){[$\mathbf{S}$ \textit{in} $\alpha_{i} |s_{i} \rangle$]}

\put(1.3,7.8){$\mathcal{V}_{\mathbf{O}} (\mathit{t_{0}})$}
\put(3,7){\vector(0,1){1.5}}
\put(6.8,7.8){$\mathcal{V}_{\mathbf{O}} (\mathit{t_{1}})$}
\put(8.5,7){\vector(0,1){1.5}}
\put(12.3,7.8){$\mathcal{V}_{\mathbf{O}} (\mathit{t_{2}})$}
\put(13.9,7){\vector(0,1){1.5}}

\put(0.5,6.5){$|\mathbf{S} (\mathit{t_{0}})\rangle \otimes |\mathbf{E} (\mathit{t_{0}})\rangle \otimes |\mathbf{O} (\mathit{t_{0}})\rangle$}
\put(5.7,6.5){$\mapsto$}
\put(6.3,6.5){$|\mathbf{S}_{\mathit{o}} \rangle \otimes |\mathbf{E} (\mathit{t_{1}})\rangle \otimes \mathit{\sum_{i} \lambda_{i} |o_{i} \rangle}$}
\put(11.3,6.5){$\mapsto$}
\put(12,6.5){$|\mathbf{S}_{\mathit{i}} \rangle \otimes |\mathbf{E} (\mathit{t_{2}})\rangle \otimes \alpha_{i} |o_{i} \rangle$}

\put(4.5,5.9){$\mathcal{R}$(\textit{Preparation})}
\put(11,5.9){$\mathcal{R} \mathit{(M_{s})}$}

\put(.2,5.5){$\mathcal{H}_{\mathbf{S}}\mathit{(t_{0})} \otimes \mathcal{H}_{\mathbf{E}}\mathit{(t_{0})} \otimes \mathcal{H}_{\mathbf{O}}\mathit{(t_{0})}$}
\put(5.2,5.6){\vector(1,0){.8}}
\put(6.2,5.5){$\mathcal{H}_{\mathbf{S}}\mathit{(t_{1})} \otimes \mathcal{H}_{\mathbf{E}}\mathit{(t_{1})} \otimes \mathcal{H}_{\mathbf{O}}\mathit{(t_{1})}$}
\put(11.2,5.6){\vector(1,0){.8}}
\put(12.2,5.5){$\mathcal{H}_{\mathbf{S}}\mathit{(t_{2})} \otimes \mathcal{H}_{\mathbf{E}}\mathit{(t_{2})} \otimes \mathcal{H}_{\mathbf{O}}\mathit{(t_{2})}$}

\put(1.3,4.3){$\mathcal{D}_{\mathbf{SO}} (\mathit{t_{0}})$}
\put(3,3.5){\vector(0,1){1.5}}
\put(6.8,4.3){$\mathcal{D}_{\mathbf{SO}} (\mathit{t_{1}})$}
\put(8.5,3.5){\vector(0,1){1.5}}
\put(12.3,4.3){$\mathcal{D}_{\mathbf{SO}} (\mathit{t_{2}})$}
\put(13.9,3.5){\vector(0,1){1.5}}

\put(4.5,3.3){$e^{-(i/ \hbar)H_{\mathbf{U}}(t)} |_{t_{0} \rightarrow t_{1}}$}
\put(2.7,3){$\mathcal{H}_{\mathbf{U}}$}
\put(4,3.1){\vector(1,0){3.5}}
\put(8.2,3){$\mathcal{H}_{\mathbf{U}}$}
\put(9.5,3.1){\vector(1,0){3.5}}
\put(13.6,3){$\mathcal{H}_{\mathbf{U}}$}
\put(10,3.3){$e^{-(i/ \hbar)H_{\mathbf{U}}(t)} |_{t_{1} \rightarrow t_{2}}$}

\put(0.5,1.5){\textit{Fig. 3:}  Explicit representation of the information about $\mathbf{S}$ encoded by $\mathbf{O}$.  The operator}
\put(0.5,1){$\mathcal{V}_{\mathbf{O}}$ maps physical states of $\mathbf{O}$ to the information that they encode.  The initial state}
\put(0.5,0.5){$|\mathbf{O} (\mathit{t_{0}})\rangle$ encodes the \textit{classical} information `[$\mathbf{S}$]' specified by the decomposition $\mathcal{D}_{\mathbf{SO}}$.}
\end{pspicture}

The classical information [$\mathbf{S}$] imposed on the physics by the decomposition $\mathcal{D}_{\mathbf{SO}}$ is represented explicitly in Fig. 3.  This diagram therefore commutes: [$\mathbf{S}$ \textit{in} $\alpha_{i} |s_{i} \rangle$] is not a physical state, so $\mathcal{V}_{\mathbf{O}} (\mathit{t_{2}}) \circ \mathcal{D}_{\mathbf{SO}} (\mathit{t_{2}}) \circ e^{-(i/ \hbar)H_{\mathbf{U}}(t)} |_{t_{0} \rightarrow t_{2}} = \mathcal{V}_{\mathbf{O}} (\mathit{M_{s}}) \circ \mathcal{V}_{\mathbf{O}} (\mathit{Preparation}) \circ \mathcal{V}_{\mathbf{O}} (\mathit{t_{0}}) \circ \mathcal{D}_{\mathbf{SO}} (\mathit{t_{0}})$ introduces no paradoxical physical ``collapse.''  Explicitly distinguishing quantum states as physical entities from the classical information that they encode thus provides a formal resolution of the measurement problem as it is standardly conceived.  It also resolves the problem posed by the apparent time-dependence of the TPS $\mathcal{H}_{\mathbf{S}} \otimes \mathcal{H}_{\mathbf{E}} \otimes \mathcal{H}_{\mathbf{O}}$ under ordinary laboratory circumstances by recognizing that relevant classical information may be encoded by different partitions of $\mathcal{H}_{\mathbf{U}}$ at different times.  Because the partitions imposed by $\mathcal{D}_{\mathbf{SO}}$ need not be regarded as \textit{physical} partitions in this representation, no assumptions of physical separability are implied.

The formal resolution of the measurement problem enabled by an explicit representation of encoded information is, moreover, a familiar one; it is identical to the formal resolution of the problem of characterizing physical behavior as computation that is reflected in the Church-Turing thesis.  Consider, for example, the characterization of the physical behavior of a laptop computer as an execution trace of some algorithm $\mathcal{A}$.  As a physical device $\mathbf{C}$, the laptop computer is a quantum system like any other.  If its interaction with the environment (and hence with a user) is neglected, it can be described by a state $|\mathbf{C}\rangle$ in a Hilbert space $\mathcal{H}_{\mathbf{C}}$ that evolves under the action of some self-propagator $e^{-(i/ \hbar)H_{\mathbf{C}}(t)}$; this is precisely the Hamiltonian oracle view of computation \cite{farhi:96}.  Describing the laptop as a \textit{classical} computer requires imposing a decomposition $\mathcal{D}_{\mathbf{C}}$ on $\mathcal{H}_{\mathbf{C}}$ that identifies particular physical degrees of freedom, e.g. particular voltage levels, with particular spatially-bounded and temporally-stable macroscopic or at least mesoscopic components, e.g. particular transistors.  Multiple such decompositions are possible, and they have no effect on $H_{\mathbf{C}}$, i.e. decompositional equivalence is satisfied.  The interpretation of the behavior of $\mathbf{C}$ as an execution trace of an algorithm $\mathcal{A}$ acting on a defined classical data structure $\mathbf{D}$ is then accomplished via a diagram such as Fig. 4, which for simplicity illustrates only three time steps, and treats the generally multi-layered semantic mapping from decompositionally-specified ``hardware'' states to abstract states of $\mathbf{D}$ as a single semantics $\mathcal{V}_{\mathbf{C}}$.  This diagram is clearly structurally equivalent to Fig. 3, and it commutes without the introduction of any physical paradox.

\psset{xunit=1cm,yunit=1cm}
\begin{pspicture}(0,0)(16,10.5)

\put(5,9.4){$\mathcal{A}_{\mathit{1}}$}
\put(11,9.4){$\mathcal{A}_{\mathit{2}}$}

\put(2.7,9){[$\mathbf{D}_{\mathit{1}}$]}
\psline{->>}(4,9.1)(7.5,9.1)
\put(8.1,9){[$\mathbf{D}_{\mathit{2}}$]}
\psline{->>}(9.5,9.1)(13,9.1)
\put(13.5,9){[$\mathbf{D}_{\mathit{3}}$]}

\put(1.3,7.8){$\mathcal{V}_{\mathbf{C}} (\mathit{t_{0}})$}
\put(3,7){\vector(0,1){1.5}}
\put(6.8,7.8){$\mathcal{V}_{\mathbf{C}} (\mathit{t_{1}})$}
\put(8.5,7.6){\vector(0,1){.8}}
\put(12.3,7.8){$\mathcal{V}_{\mathbf{C}} (\mathit{t_{2}})$}
\put(13.9,7){\vector(0,1){1.5}}

\put(0,6.5){$|\mathbf{T1} (\mathit{t_{0}})\rangle \otimes |\mathbf{T2} (\mathit{t_{0}})\rangle \otimes ... \otimes |\mathbf{TN} (\mathit{t_{0}})\rangle$}
\put(6.7,6.5){$\mapsto$}
\put(7.6,7){$|\mathbf{T1} (\mathit{t_{1}})\rangle \otimes$}
\put(10,6.5){$\mapsto$}
\put(10.7,6.5){$|\mathbf{T1} (\mathit{t_{2}})\rangle \otimes |\mathbf{T2} (\mathit{t_{2}})\rangle \otimes ... \otimes |\mathbf{TN} (\mathit{t_{2}})\rangle$}

\put(7.1,6.5){$\Bigg\{$}
\put(7.4,6.5){$|\mathbf{T2} (\mathit{t_{1}})\rangle \otimes ...$}
\put(7.6,6){$\otimes |\mathbf{TN} (\mathit{t_{1}})\rangle$}
\put(9.6,6.5){$\Bigg\}$}

\put(.4,5.5){$\mathcal{H}_{\mathbf{T1}} \otimes \mathcal{H}_{\mathbf{T2}} \otimes ... \otimes \mathcal{H}_{\mathbf{TN}} (\mathit{t_{0}})$}
\put(5.5,5.5){$\longrightarrow$}
\put(6.2,5.5){$\mathcal{H}_{\mathbf{T1}} \otimes \mathcal{H}_{\mathbf{T2}} \otimes ... \otimes \mathcal{H}_{\mathbf{TN}} (\mathit{t_{1}})$}
\put(11.3,5.5){$\longrightarrow$}
\put(12,5.5){$\mathcal{H}_{\mathbf{T1}} \otimes \mathcal{H}_{\mathbf{T2}} \otimes ... \otimes \mathcal{H}_{\mathbf{TN}} (\mathit{t_{2}})$}

\put(3.8,5){$ $}
\put(9.5,5){$ $}

\put(1.3,4.3){$\mathcal{D}_{\mathbf{C}} (\mathit{t_{0}})$}
\put(3,3.5){\vector(0,1){1.5}}
\put(6.8,4.3){$\mathcal{D}_{\mathbf{C}} (\mathit{t_{1}})$}
\put(8.5,3.5){\vector(0,1){1.5}}
\put(12.3,4.3){$\mathcal{D}_{\mathbf{C}} (\mathit{t_{2}})$}
\put(13.9,3.5){\vector(0,1){1.5}}

\put(4.5,3.3){$e^{-(i/ \hbar)H_{\mathbf{C}}(t)} |_{t_{0} \rightarrow t_{1}}$}
\put(2.7,3){$\mathcal{H}_{\mathbf{C}}$}
\put(4,3.1){\vector(1,0){3.5}}
\put(8.2,3){$\mathcal{H}_{\mathbf{C}}$}
\put(9.5,3.1){\vector(1,0){3.5}}
\put(13.6,3){$\mathcal{H}_{\mathbf{C}}$}
\put(10,3.3){$e^{-(i/ \hbar)H_{\mathbf{C}}(t)} |_{t_{1} \rightarrow t_{2}}$}

\put(0.5,2){\textit{Fig. 4:}  Interpretation of the behavior of a physical device $\mathbf{C}$ as an execution trace of}
\put(0.5,1.5){an algorithm $\mathcal{A}$ on a classical data structure $\mathbf{D}$.  The decomposition operator $\mathcal{D}_{\mathbf{C}}$ that}
\put(0.5,1){defines the ``hardware'' states $|\mathbf{T1}\rangle$ ... $|\mathbf{TN}\rangle$ has no effect on the physical dynamics $H_{\mathbf{C}}$.}
\put(0.5,0.5){Time dependence of TPS components has been abbreviated to preserve the layout.}
\end{pspicture}

It is important to emphasize that Fig. 3 does not reflect a non-ontic view of quantum states, just as Fig. 4 does not reflect a non-ontic view of laptop computers.  Indeed it reflects precisely the opposite view: quantum states are physical states, but all \textit{classical} states are virtual.  This conclusion follows directly from decompositional equivalence.  Because decompositions of $\mathcal{H}_{\mathbf{U}}$ into TPSs have no effects on $H_{\mathbf{U}}$, they can be interchanged without changing anything physical.  The category $(\mathcal{H}, \mathit{A}, \otimes)$ can, therefore, have no physical meaning beyond $(\mathcal{H}, \mathit{A})$; the tensor-product operator $\otimes$ has significance for the semantics, but no physical significance.  The decomposition operator $\mathcal{D}_{\mathbf{SO}}$ and the problematic ``states'' that it generates can, therefore, be dropped from Fig. 3 altogether to produce the much simpler representation shown in Fig. 5, in which the semantics $\mathcal{V}_{\mathbf{O}}$ is taken to act directly on $|\mathbf{U}\rangle$.

\psset{xunit=1cm,yunit=1cm}
\begin{pspicture}(0,0)(16,7)

\put(3.5,5.9){$\mathcal{V}_{\mathbf{O}}$(\textit{Preparation})}
\put(11,5.9){$\mathcal{V}_{\mathbf{O}}(\mathit{M_{s}})$}

\put(2.7,5.5){[$\mathbf{S}$]}
\psline{->>}(3.5,5.6)(6,5.6)
\put(6.5,5.5){[$\mathbf{S}$ \textit{prepared in} $\{ |s_{i} \rangle \}$]}
\psline{->>}(10.5,5.6)(12.5,5.6)
\put(13,5.5){[$\mathbf{S}$ \textit{in} $\alpha_{i} |s_{i} \rangle$]}

\put(1.3,4.3){$\mathcal{V}_{\mathbf{O}} (\mathit{t_{0}})$}
\put(3,3.5){\vector(0,1){1.5}}
\put(6.8,4.3){$\mathcal{V}_{\mathbf{O}} (\mathit{t_{1}})$}
\put(8.5,3.5){\vector(0,1){1.5}}
\put(12.3,4.3){$\mathcal{V}_{\mathbf{O}} (\mathit{t_{2}})$}
\put(13.9,3.5){\vector(0,1){1.5}}

\put(4.5,3.3){$e^{-(i/ \hbar)H_{\mathbf{U}}(t)} |_{t_{0} \rightarrow t_{1}}$}
\put(2.5,3){$|\mathbf{U} (\mathit{t_{0}})\rangle$}
\put(4,3.1){\vector(1,0){3.5}}
\put(4,3){\line(0,1){.2}}
\put(8,3){$|\mathbf{U} (\mathit{t_{1}})\rangle$}
\put(9.5,3.1){\vector(1,0){3.5}}
\put(9.5,3){\line(0,1){.2}}
\put(13.4,3){$|\mathbf{U} (\mathit{t_{2}})\rangle$}
\put(10,3.3){$e^{-(i/ \hbar)H_{\mathbf{U}}(t)} |_{t_{1} \rightarrow t_{2}}$}

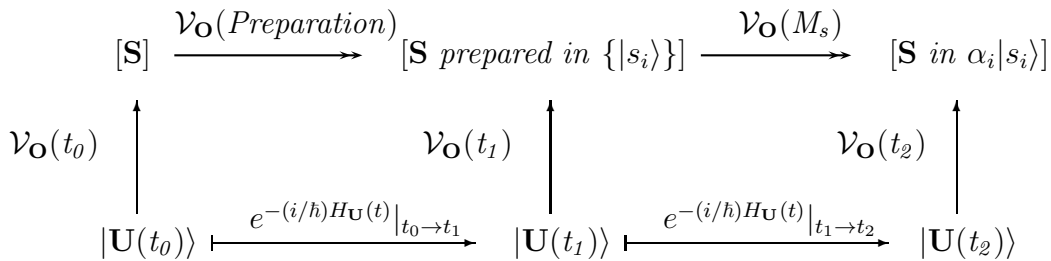
\put(0.5,1.5){\textit{Fig. 5:}  Simplification of Fig. 3 recognizing that decomposition into a TPS has no}
\put(0.5,1){physical consequences and hence that the semantics $\mathcal{V}_{\mathbf{O}}$ can be regarded as acting}
\put(0.5,0.5){directly on the physical state $|\mathbf{U}\rangle$.}
\end{pspicture}

The diagram shown in Fig. 5 cleanly raises the question at the core of the quantum measurement problem: what is the physical implementation of the semantics $\mathcal{V}_{\mathbf{O}}$?  What does it mean to say that one collection of degrees of freedom of $\mathbf{U}$ encodes classical information about another collection of degrees of freedom of $\mathbf{U}$, and how can such an encoding, once established, remain stable over time?  As suggested in \cite{fields:12c}, the only available answer to the former question appears to be entanglement, a means of generating recordable classical correlations that is itself decomposition- and observable-dependent \cite{zanardi:01, zanardi:04, torre:10, harshman:11, thirring:11, lychkovskiy:13} and hence non-objective \cite{bartlett:07}.  The answer to the latter would appear to depend entirely on the details of $H_{\mathbf{U}}$, details that at least appear not to generalize across encodings.

\section{Conclusion}

Viewing measurement as a physical process in which the system of interest acts on the observer serves to emphasize a conclusion derivable from decompositional equivalence alone: if minimal quantum theory is assumed to be correct, classical states can only consistently be described as virtual.  Every classical ``it,'' in other words, is a ``bit'' and nothing more.  Recognizing this provides a formal resolution to the measurement problem as standardly formulated, one that is familiar from the classical semantics of programming languages.  This formal resolution raises the deeper question of how $H_{\mathbf{U}}$ physically implements a classical semantics.  The demonstrated decomposition-dependence of entanglement suggests that this question has at best a relational answer.

\end{document}